# A COMPACT AND SUSTAINABLE ELECTRONIC MODULE FOR SILICON PHOTOMULTIPLIERS

A. Sadigov[1,2], S. Nuruyev[2,3], R. Akbarov[1,2,3], D.B. Berikov[3,4], A. Madadzada[1,3], A. Mammadli[2]

[1]Innovation and Digital Development Agency Nuclear Research Department, Baku, Azerbaijan
[2]Institute of Radiation Problems under Ministry of Science and Education, Baku, Azerbaijan
[3]Joint Institute for Nuclear Research, Dubna, Russia
[4]Institute of Nuclear Physics of the Ministry of Energy of the Republic of Kazakhstan, Almaty, Kazakhstan
e-mail: sebuhinuruyev@jinr.ru

**Abstract**

This article presents the development of a cost-effective and efficient electronic module for silicon photomultipliers (SiPM). The electronic module combines essential functionalities, such as a high voltage power supply, a preamplifier, and a signal comparator, into a compact circuit. A high voltage power supply with a range of 30 to 140 V provides a stable bias voltage with 0.01 V accuracy for the SiPMs, while a preamplifier with 40 gain and 250 MHz bandwidth enables signal amplification necessary to extract weak signals. The comparator converts an analogue signal (higher than 8 mV) into TTL (transistor-transistor logic), which makes it easy to process and analyze with digital devices such as microcontrollers or make it possible to send signals over long distances by a cable. The module has been tested using an LYSO scintillator and a SiPM called a micropixel avalanche photodiode (MAPD). It provides a more effective and efficient solution for reading out signals from SiPMs in a variety of applications, delivering reliable and accurate results in real-time.

## 1. Introduction

Nowadays, silicon photomultipliers (SiPMs) have emerged as solid-state photodetectors with immense potential, serving as compelling alternatives to vacuum photomultipliers (PMTs) across a broad spectrum of applications [1-4]. SiPMs come with a multitude of advantages, notably their robustness, fast response times, low operating voltage requirements, and compactness [5-8]. These attributes have propelled the widespread adoption of SiPMs in diverse fields, encompassing particle physics, astrophysics, nuclear physics, medicine, and public security applications [9-15].

Within high-energy particle experiments, SiPMs are indispensable devices for the detection of various types of particles [16]. They also find utility in specialized detectors designed for the observation of faint astronomical phenomena, such as gamma-ray bursts and supernovae [17]. In the field of nuclear physics, SiPMs contribute to the precise measurement of reactor antineutrino spectra [18]. Furthermore, SiPMs play a pivotal role in medical diagnostics, particularly in Positron Emission Tomography (PET) imaging, obtaining detailed body images with remarkable precision [19].

While these applications frequently employ Specific Integrated Circuits (ASICs) to handle the substantial number of output channels generated by SiPM arrays, they prove impractical for experiments with low channel density requirements. To address this challenge, a compact and sustainable electronic module has been developed specifically for SiPM readout systems. This module integrates functionalities, including bias supply, discriminator, and amplifier, into a single circuit. Consequently, it simplifies the overall circuitry and enhances the usability of SiPMs in various experiments.

This article aims to provide the design and performance of this compact and sustainable electronic module tailored for readout systems employing silicon photomultipliers.

## 2. Electronic module

The module is a versatile device designed to cater to various applications where different SiPMs are used. The aim of this section is to shed light on possibilities of the module through a comprehensive analysis of its parts.

This module encompasses a DC-DC voltage converter, a preamplifier, and a comparator that convert analogue signal to TTL. Each component of the module plays a crucial role in enhancing functionality and performance. In this article, we delve into the intricate details of each component, examining their functionalities, design principles, and performance characteristics.

## 2.1. The high voltage converter

The high voltage converter is very stable and the average fluctuation of output voltage of the DC-DC converter is less than 0.01 V. Figure 1 shows the circuit diagram and the final form of the assembled DC-DC circuit.

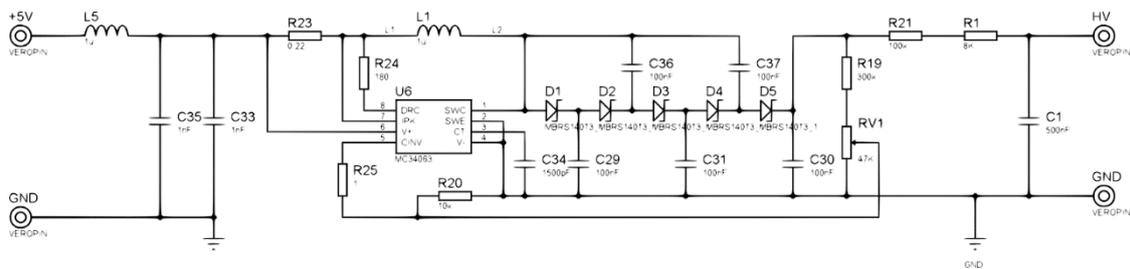

Figure 1. The high voltage converter circuit diagram.

The voltage controller MC34063A was used as a pulse generator. Low-capacity capacitors (30 pF) were used to control the frequency of the generated signal. At this time, the frequency of the rectangular or sinusoidal signal was 10500 Hz. When choosing this value of the frequency, the current demand of the controller was minimum. Then the signal is fed to the coil with an inductance of 1000 μH and the energy is collected in the inductance and the capacitor of 100 nF is discharged by means of the BAV.102 diode depending on the frequency. By increasing the inductance charging time, it is possible to increase the output voltage. The energy collected in the inductance is discharged to the capacitors by means of cascades consisting of a pair of diodes and capacitors, and when the charged capacitors are discharged, the output voltage increases as much as the voltage collected in the capacitor. In this way, it is possible to increase the output voltage by increasing the number of capacitors and diode pairs at a given frequency. An additional voltage divider is assembled to change (decrease and increase) the applied voltage in a voltage interval of 10 V. Zenner diodes were used to ensure the stability of the output voltage. In addition, the feedback channel of the MC34063A was utilized to ensure that the output voltage remains independent of the input voltage. This ensured a constant output value even when the input signal varies within the range of 3-10 V.

The IC circuit was initially simulated using ISIS 7 Professional Proteus software and later assembled on the board. Special attention was given to minimizing current loss during the selection of components. The current demand of this type of voltage converter taken from the input was 2 mA [20].

## 2.2. The preamplifier

The preamplifier for SiPMs with NPN and PNP transistors is designed to convert the weak photocurrent signals from SiPMs and SiPM based detectors into amplified voltage signals. It provides a gain of 40 and a bandwidth of 250 MHz, allowing for accurate and efficient signal amplification suitable for silicon photomultiplier applications. This was necessary because the duration of scintillation photons emitted from all the type scintillators with decay time more than 20 ns. Additionally, the preamplifier can increase its bandwidth up to 500 MHz with minimal adjustments. Figure 2 shows the circuit diagram and the final form of the assembled preamplifier.

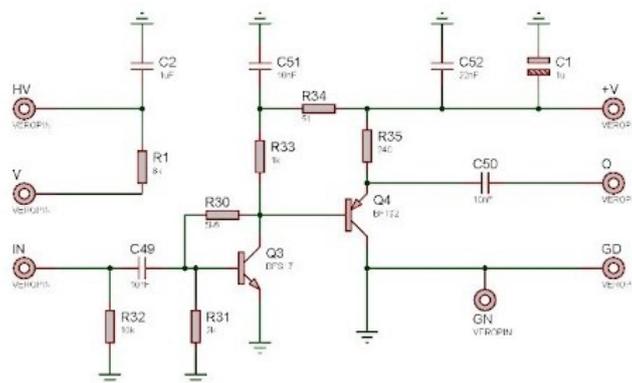

Figure 2. The circuit diagram of the preamplifier.

## 2.3. The comparator

Signal comparator was implemented to convert the negative analogue output signal from the preamplifier to TTL levels. An analogue-TTL converter was assembled using an LT-1355 operational amplifier. Figure 3 shows the circuit diagram and the final form of the comparator.

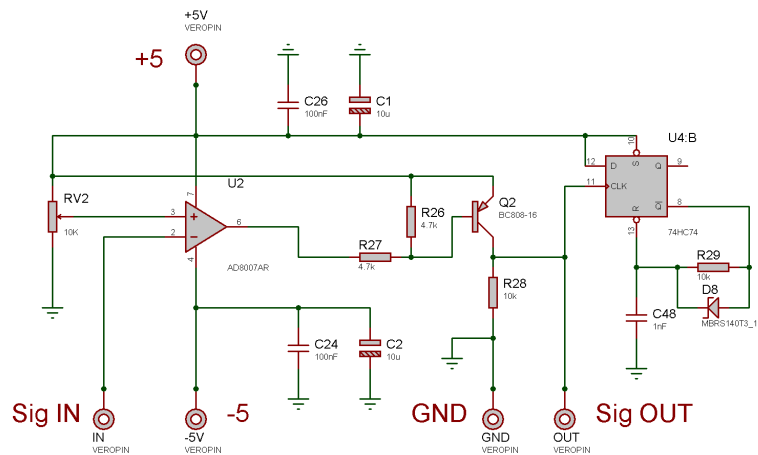

Figure 3. The circuit diagram of the comparator.

The analogue signal received from the preamplifier was connected to the input of the comparator. The threshold level for the input signal was determined using a pot resistor. If the input signal exceeded the threshold value (set at 8 mV), input signal was compared to the reference one, resulting in a square signal at the output. When the input signal amplitude was below the threshold, no signal was observed at the output, which was held at a logic low state (0). As the input signal amplitude increased, changes in the output signal amplitude of the comparator were observed. To maintain a constant amplitude for the output signals, the signal received at the output of the comparator was fed to the base of a transistor. This arrangement ensured that the output signals had a consistent amplitude level. To prevent changes in the signal width, a 74HC74 flip-flop was added to the circuit. This flip-flop helped maintain a constant width for the signal received at the digital output. Time shift between two inputs of the flip-flop defines the widths of the output signal.

**3. Electronic module testing**

The detector signal based on the MAPD and LYSO scintillator was used to check the functionality of the module. The scintillator was wrapped with several layers of white Teflon tape on all sides (to reduce light loss), except for the side attached to the MAPD-3NM-II [21]. For better optical contact between the scintillator and MAPD, a special optical lubricant was used. The $^{137}$Cs calibration source was placed on the surface of the LYSO scintillator. A CAEN DT5720 multichannel desktop signal digitizer with a sampling rate of 250 MHz was used as an analysis and recording device. The measurements were carried out in a so-called dark box, to isolate the detector from ambient light, at room temperature (Fig. 4).

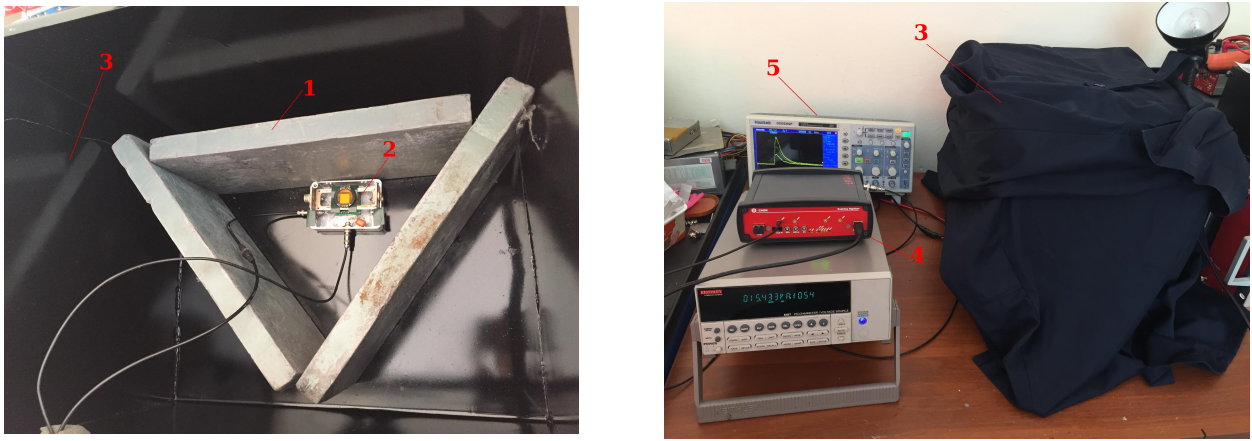

Figure 4. Setup for testing the electronic module. 1 - lead shield to reduce the background, 2 - detector connected to the module, 3 - dark box, 4 - desktop digitizer CAEN, 5 – oscilloscope.

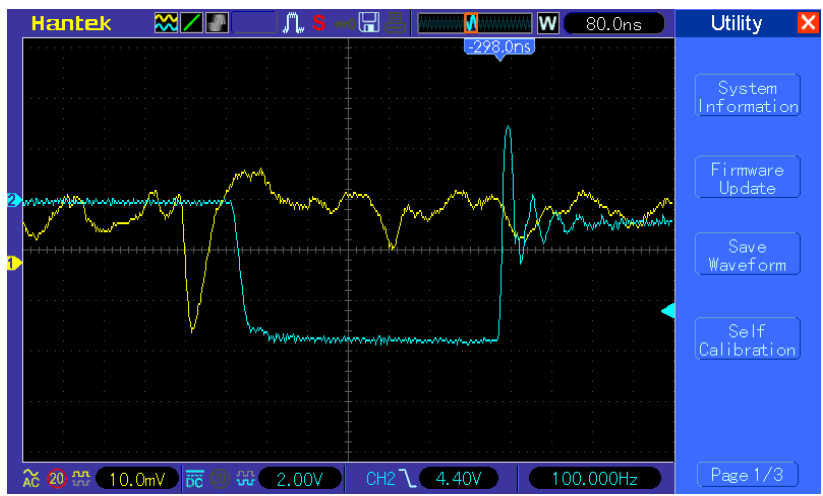

Figure 5. The output waveform from the module, measured with an oscilloscope. The digital pulse (blue) was converted from analogue (yellow) to TTL (inverted) standard for counting and triggering purposes.

The output signal of the module branched: one output was fed to an oscilloscope for visual observation of pulses (Fig. 5), the other to a digitizer. Then the digitized signal was sent to a personal computer for data accumulation and their subsequent analysis. Subsequent analysis of the waveform data was carried out using scripts written in the ROOT environment.

The spectrum recorded with the $^{137}$Cs gamma source is presented in Figure 6. The energy resolution for the 662 keV $^{137}$Cs gamma yield was quantified at 10±0.3% for the detector based on MAPD and LYSO scintillator [22].

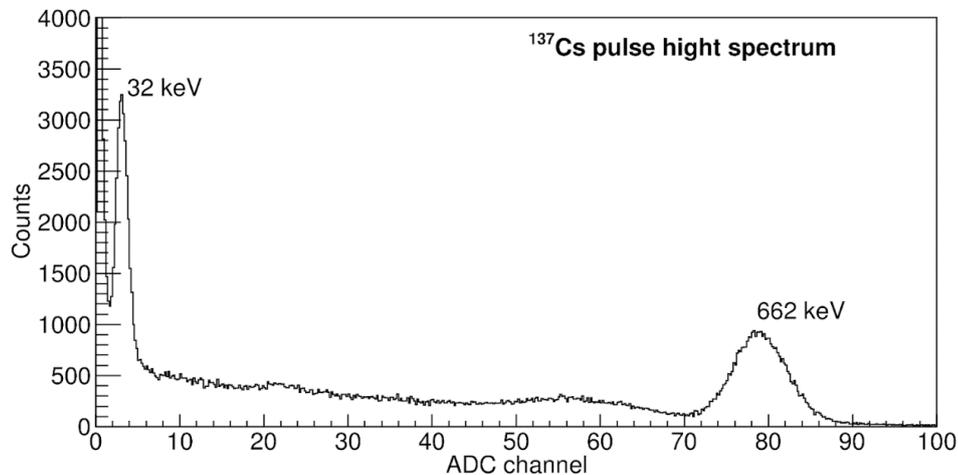

Figure 6. $^{137}$Cs spectra on count and energy mode.

This level of precision aligns with the registration of the 32 keV gamma signature originating from $^{137}$Cs by the module. These observations collectively imply a notable reduction in electronic noise within the module. Beyond its economic viability, the module offers a combination of cost-effectiveness, compactness, and portability. Despite its economical construction, when scrutinized through a scientific lens, the module yields results of commendable quality that duly fulfill requisites.

**Conclusion**

We have developed a cost-effective and efficient electronic module that integrates multiple electronic components, including a DC-DC step-up voltage converter, a preamplifier, and a comparator, into a compact circuit. It provides a bias voltage in the range of 30 V to 140 V with accuracy of 0.01 V. The preamplifier linearly amplifies signals received from most scintillators commonly used in industry, medicine, and experiments, with decay times ranging from 30 to 100 ns. This module provides a sustainable and economical solution for detectors based on silicon photomultipliers and scintillators for ionization radiation detection, particularly for energy determinations and counting purposes. The module has been tested with a detector based on MAPD and LYSO scintillator using a Cs-137 gamma source. The energy resolution achieved was 10±0.3% for 662 keV gamma rays. Counting performance of the module is approximately 2·$10^6$, considering a TTL output signal width of 450 ns. It is possible to reduce the signal width with minor modifications. The portability and compact design of this developed module make it a cost-effective and efficient solution that combines multiple functions. This electronic module can become an essential tool in experimental setups requiring compact and stable electronic circuits.